\documentclass[journal=jacsat,manuscript=article]{achemso}

\usepackage[version=3]{mhchem}
\usepackage{graphicx}
\usepackage{dcolumn}
\usepackage{bm}
\usepackage{latexsym}
\usepackage{color}
\usepackage{hyperref}
\usepackage{amsmath}
\usepackage{amssymb}
\usepackage{natbib}
\usepackage{multicol}
\SectionNumbersOn
\author{Evan L. H. Thomas}
\altaffiliation{These authors contributed equally to the work.}
\email{elhthomas@gmail.com}
\author{Soumen Mandal}
\altaffiliation{These authors contributed equally to the work.}
\email{soumen.mandal@gmail.com} \affiliation[Physics and
Astronomy]{School of Physics and Astronomy, Cardiff University,
Cardiff, UK}
\author{Emmanuel B. Brousseau}
\affiliation[Engineering]{Cardiff School of Engineering, Cardiff
University,
Cardiff, UK}%
\author{Oliver A. Williams}
\email{willamso@cf.ac.uk} \affiliation[Physics and
Astronomy]{School of Physics and Astronomy, Cardiff University,
Cardiff, UK}

\title{Polishing of \{100\} and \{111\} single crystal diamond through the use of Chemical Mechanical Polishing} 
\begin{document}

\begin{abstract}
Diamond is one of the hardest and most difficult to polish
materials. In this paper, the polishing of \{111\} and \{100\}
single crystal diamond surfaces by standard Chemical Mechanical
Polishing, as used in the silicon industry, is demonstrated. A
Logitech Tribo Chemical Mechanical Polishing system with Logitech
SF1 Syton and a polyurethane/polyester polishing pad was used. A
reduction in roughness from 0.92 to 0.23 nm root mean square (RMS)
and 0.31 to 0.09 nm RMS for \{100\} and \{111\} samples
respectively was observed.
\end{abstract}

\section{\label{sec:intro}Introduction}
Diamond has long been used for cutting and polishing applications
due to its extreme hardness, high thermal conductivity and
chemical inertness.  However, properties such as a large band gap,
high resistivity, high electron mobility, low dielectric constant,
and low thermal coefficient of expansion make diamond an excellent
candidate for high power - high frequency
electronics\cite{Field1992, Kohn2001}, and optical
devices\cite{Schuelke2013, Teraji2004}. With the advancement of
technology it is now possible to economically synthesize high
quality large area single crystal diamonds through
homoepitaxial CVD\cite{Isberg2002, Tallaire2005}. However, to
prevent defects and surface damage from the substrate from
propagating into the CVD layer a polishing step is
required\cite{Friel2009}. Recent experiments have also pointed to
the presence of two dimensional hole gas on the surface of
hydrogen terminated diamond\cite{Hauf2013}. To harness this
phenomenon, and for the applications mentioned above, it is also
important to have atomically flat, defect free top surfaces -
necessitating an efficient polishing technique.

For the polishing of diamond, while many techniques exist
including thermo-mechanical, ion beam, and thermal annealing,
mechanical polishing has traditionally prevailed\cite{Malshe1999,
Jin1992, Suzuki1996, Hirata1992, Ozkan1997, Olsen2004}. Mechanical
polishing is typically done through the use of a fast rotating
metal scaife charged with a diamond grit and olive oil binder. The
sample is polished by putting it under pressures of 2.5-6.5 MPa
for grinding and 1-2.5 MPa for polishing against a scaife with
linear velocity of approximately 50 m/s\cite{Chen2013}. However,
the polishing of diamond is highly anisotropic with two orders of
magnitude difference in removal rate between hard and soft
polishing directions for the \{100\} and \{110\} plane
groups\cite{Grillo1997}. Along soft directions polishing is the
result of shearing between diamond chips on the scaife and the
sample surface, driving a phase conversion to non sp$^3$
material\cite{Pastewka2011}. As a result `nano-grooves' are found
on the surface with lengths of 20 - 1000 nm dependent on the grit used and depths of up to 20 nm, whereas, polishing along hard
directions leads to fracturing along the \{111\} plane producing a
rough `hill and valley' type surface\cite{Hird2004}.  Polishing of
the \{111\} plane meanwhile remains difficult\cite{Huisman1997},
with only slight anisotropy between hard and softer polishing
direction\cite{Schuelke2013}. Due to the inherent mechanical
nature of this technique subsurface damage occurs, with fractures
at the surface propagating into the bulk\cite{Volpe2009}. This
problem of poor surface quality is often seen in the techniques
mentioned above, preventing full use of the properties of
diamond\cite{Gaisinskaya2009}.

In order to reduce these polishing artifacts several methods have
been proposed as a finishing technique, including
chemo-mechanical\cite{Kuhnle1995, Kubota2012, Wang2006} or
tribochemical polishing\cite{Haisma1992}, and reactive ion
etching\cite{Volpe2009}. In chemo-mechanical polishing a molten
oxidizer is added, typically KNO$_3$ (potassium nitrate), NaNO$_3$
(sodium nitrate)\cite{Kuhnle1995}, or H$_2$O$_2$ (hydrogen
peroxide)\cite{Kubota2012}. Between the scaife and diamond sample
hot spots of 360 $^o$C are reached driving a conversion to CO and
CO$_2$\cite{Kuhnle1995}. While this method achieves low roughness
values, the use of scaife and elevated temperatures makes the
process complicated and very different to that used in the silicon
based electronics industry\cite{Krishnan2009}.

Previous work by the authors have shown Chemical Mechanical
Polishing, a technique used in the polishing of gate dielectrics
in IC fabrication, can successfully be used to polish
Nanocrystalline Diamond (NCD) films\cite{Thomas2014}.  In this
technique the sample is swept across the polyester/polyurethane
based polishing pad doused with colloidal silica polishing fluid
(Syton) without the use of any diamond grit or elevated
temperatures.  Drawing parallels with the mechanism used to
describe the polishing of SiO$_2$\cite{Hocheng2001}, it was
tentatively suggested that wet oxidation of the diamond increases
the hydroxide content on the surface, facilitating the binding of
silica particles within the slurry. Should an asperity on the
rough pad then create a sufficient shearing force on the silica
particle, the particle and attached carbon atom will be removed,
polishing the surface.

In this article the use of CMP on \{100\} and \{111\} single
crystal diamond is demonstrated through the use of Atomic Force Microscopy (AFM). The aspects
of this adaption from the IC fabrication industry, including the
condition of the polishing pad and post polishing cleaning is
also been discussed.

\section{Experimental Method}
HPHT Single Crystal \{100\} and \{111\} samples were obtained from
Elementsix. The \{100\} sample was approximately 2 by 2 by 0.5 mm
high, whereas the \{111\} sample was 3 by 3 by 0.7 mm high. Before
use both samples were given a standard SC-1 clean of 30\%
H$_2$O$_2$: NH$_4$OH:DI H$_2$O (1:1:5) at 75 $^o$C for 10 min,
followed by a ultrasonic DI H$_2$O bath for 10 minutes.  In
preparation for polishing, samples were bonded within a slight
recess on a 2-inch polymer holder with cyanoacrylate.  The recess
was then filled up with Crystalbond to prevent shear forces on
crystal edges, ensuring a stable mounting while leaving only the
surface to be polished protruding.  This template was then placed
inside a carrier chuck suitable for use with the CMP equipment.

Chemical Mechanical Polishing was carried out with a Logitech
Tribo CMP tool equipped with a SUBA - X polyester/polyurethane pad
and Syton colloidal silica alkaline polishing fluid (15-50\%
SiO$_2$, 9.2-10.1 pH, 4-5\% ethylene glycol), as described
elsewhere\cite{Thomas2014}. Before, and during use, the polishing
pad was conditioned to maintain a rough surface to maximize
polishing action and to efficiently distribute polishing slurry.
During polishing both holder and pad were kept rotating at 60 rpm
in opposite directions as the holder swept across the pad.  Down
pressure on the holder was maintained at 4 psi while a backing
pressure of 20 psi was applied to prevent bowing of the holder and
ensure contact between the diamond crystal and the polishing pad.
After initial wetting of the plate, the polishing slurry rate was
maintained at 40 ml/min.  Polishing durations for the \{100\} and
\{111\} single crystals were 3 and 7 hours respectively.  After
polishing, the samples were cleaned with SC-1 and Hydrofluoric
Acid in an attempt to remove any organic contaminants and
remaining silica.

Atomic Force Microscopy was performed with a Park Systems XE-100
AFM operating in non-contact mode equipped with NT-MDT NSG30 tips
(320 kHz resonant frequency, 40 N/m spring constant, 10 nm tip
radius).  Multiple areas of 5 $\times$ 5 $\mu$m$^2$ were scanned for
each sample before and after polishing, with analysis of data
being carried out by WSxM and Gwyddion SPM analysis software.

For analysis of the polishing pad, samples were taken of: a fresh
pad, a conditioned pad, and a pad that had been subjected to 7
hours of single crystal polishing. Scanning Electron Microscopy
(SEM) images were taken with the SE2 detector of a Raith E-line
SEM, operating at 10 kV accelerator voltage and 9 mm working
distance.

\section{Results and Discussions}
\subsection{\{100\} Single Crystal}
\begin{figure}[t]
\centerline{\includegraphics[width=14cm,angle=0]{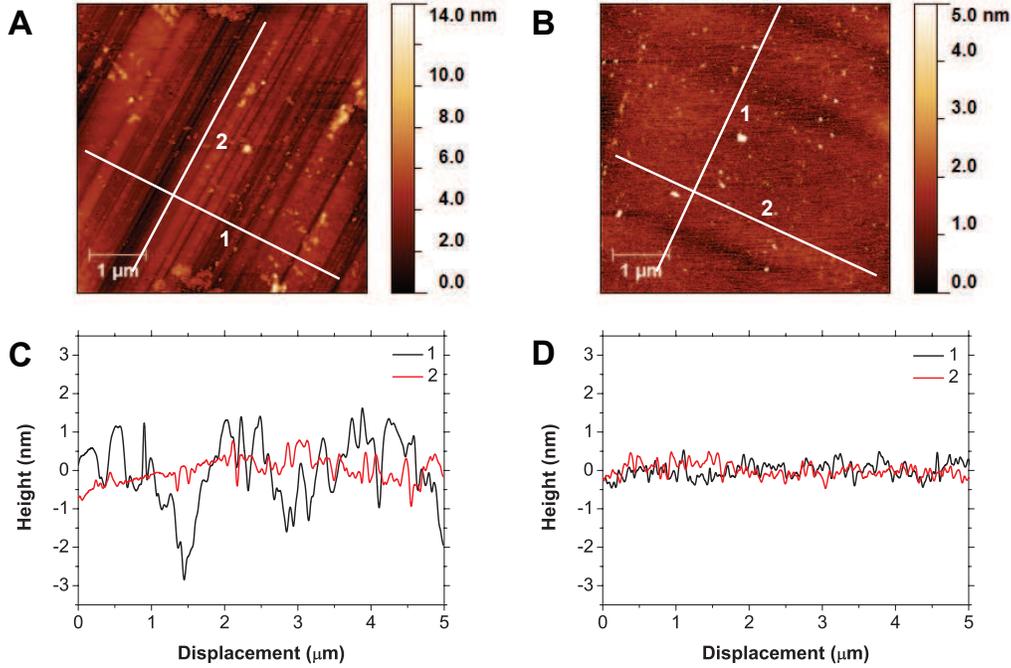}}
\caption{AFM images of a \{100\} orientated single crystal before
(A) and after chemical mechanical polishing (B).  Shown in panels
C and D are line traces perpendicular to the mechanical polishing
direction (1) and along the polishing direction (2) of the
respective AFM images.  Clear removal of the mechanically induced
phase transformation nano-grooves can be seen in the AFM images,
backed up by the dramatic reduction in amplitude of the
perpendicular to polishing direction line trace.} \label{100}
\end{figure}
Typical AFM scans of the \{100\} single crystal before and after
polishing are shown in panels A and B of Fig. \ref{100}, while
lines traces perpendicular to polishing direction (1), and along
the polishing direction (2) are plotted in panels C and D.  As can
be seen in panels A and C, the surface of the sample prior to CMP
consists of clearly defined ``nano-grooves'' as a result of the
nanodiamond particle induced phase transformation along the
$<$100$>$ soft direction. From the 5 $\mu$m perpendicular line
trace it can be seen that the nano-grooves widths are between 100 and
500 nm and that they have an average depth of 3 nm.  Meanwhile, for the trace
along the polishing direction less variation is seen with little
indication of fracture.  Roughness values are 0.92 nm RMS and 0.34
nm RMS for the perpendicular to and along polishing direction line
traces respectively.

Looking at the 3 hour CMP polished AFM image and line traces of
Fig. \ref{100}B and Fig. \ref{100}D a clear difference can be seen
with the removal of the mechanical polishing scaife marks to the
point at which it is difficult to resolve the original polishing
direction. The line trace perpendicular to the polishing direction
shows a decrease in the larger undulations of the nano-grooves to
a point at which it is in close agreement with the trace taken
along the original polishing direction, with roughness values
being 0.23 and 0.19 nm RMS respectively. It is also worth noting
that a lack of polishing debris can be seen in the polished AFM
image, showing the ease of removal of the polishing slurry used.

\subsection{\{111\} Single Crystal}
\begin{figure}[t]
\centerline{\includegraphics[width=14cm,angle=0]{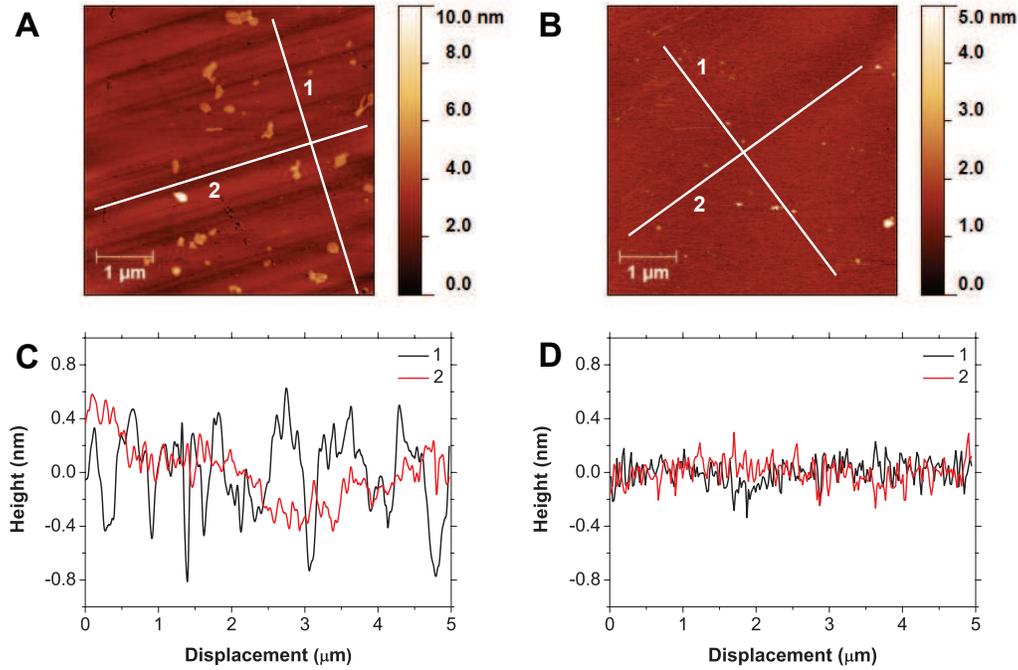}}
\caption{AFM images of the \{111\} polished plane before (A) and
after chemical mechanical polishing (B).  Shown in panels C and D
are line traces similar in fashion to those seen in Fig.
\ref{100}. Once again, a clear removal of grooved surface features
can be seen, leading to a very smooth surface with line trace
roughness reducing from 0.31 and 0.23 nm RMS for perpendicular to
and along original polishing direction to 0.09 nm RMS.}
\label{111}
\end{figure}
As with the \{100\} single crystal sample, Fig. \ref{111} shows
typical AFM images of the \{111\} single crystal before and after
being subjected to 7 hours of CMP along with line traces
perpendicular (1), and along the mechanical polishing direction
(2).  As can be seen from panel A, the \{111\} single crystal has
similar, but shallower, grooved features as the \{100\} crystal.
Roughness values seen perpendicular to and along the original
polishing direction are 0.31 nm and 0.23 nm RMS respectively.
After being subjected to CMP it is again very difficult to
determine the original polishing direction due to the dramatic
reduction in the grooved features.  Looking at the two
perpendicular line traces they appear very similar, reiterating
the removal of these grooves.  Roughness values for these traces
are significantly lower at 0.09 nm RMS for both perpendicular and
along the polishing direction traces.

\subsection{Polishing Pad}
\begin{figure}[t]
\centerline{\includegraphics[width=14cm,angle=0]{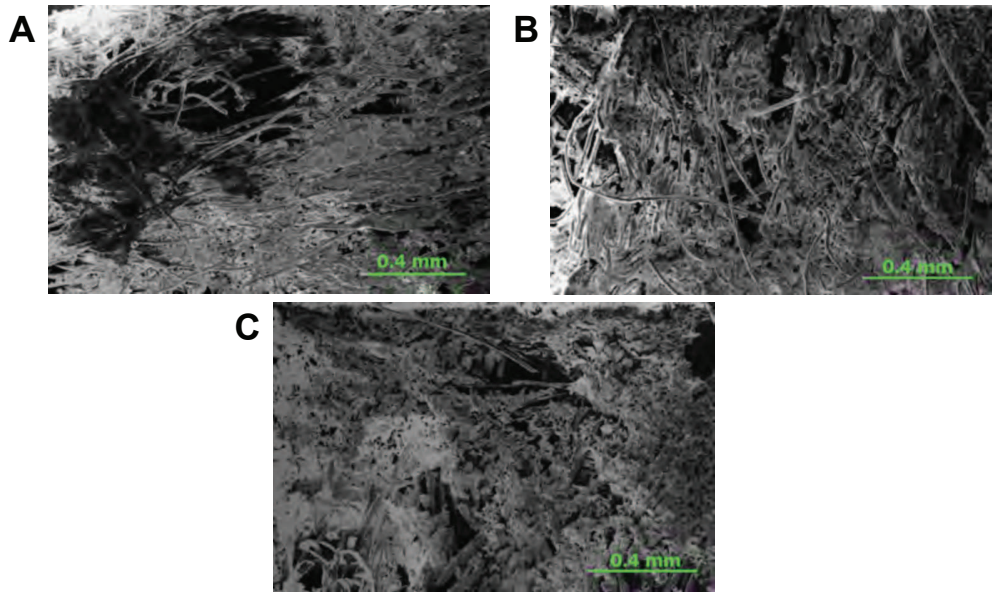}}
\caption{Polishing pad samples taken: (A) before use, (B) after
conditioning for 1 hour, and (C) after 7 hours of single crystal
polishing} \label{pad}
\end{figure}
The condition and properties of the polishing pad heavily dictate
the wear rate and uniformity of the resulting diamond
surface\cite{Hooper2002, Lu2002, Mcgrath2004, Park2007}. In
traditional CMP it has been seen that the removal rate is directly
related to the surface roughness and hence number of asperities of
the polishing pad\cite{Park2007}. However, the pad also needs to
be porous to allow for slurry distribution and clearing away of
spent material. To increase this surface roughness and increase
the porosity, pads are typically `run in'\cite{Mcgrath2004} or
`conditioned' before use with an electroplated diamond grit plate.
Upon polishing the pad is then plastically deformed by the sample,
closing pores and reducing the number of these surface asperities
in as little as 10 minutes\cite{Mcgrath2004}. While the diamond
samples themselves will condition the pad surface, due to the
polishing duration being in the hours rather than
minutes\cite{Mcgrath2004}, in-situ conditioning of the pad is also
needed. Due to the difference in mechanical properties of diamond
and the materials traditionally polished with these
polyester/polyurethane pads, work is needed to characterize the
pad during diamond polishing.

SEM images of an as received SUBA-X polishing pad, a conditioned
pad, and a pad subjected to 7 hours of single crystal CMP can be
seen in Fig. \ref{pad}A to \ref{pad}C.  From the fresh pad the
polyester strands can be seen to be bound together by the dense
polyurethane foam. After being conditioned, the abrasion by the
electroplated diamond grit can be seen.  Polyester strands are
less uniformly orientated, if not cut, and the polyurethane binder
has opened up and become rougher at the magnification used.  It
can therefore be assumed that conditioning has increased the
surface asperities on a scale closer to the size of the silica
particles.  Wear tracks visible by eye were observed on the pad
after 7 hours suggesting significant abrasion is carried out by
the diamond surfaces.  The corresponding SEM image also shows this
wear with most of the polyester strands being severed and
polyurethane foam appearing severely abraded.

\subsection{Discussion}
Looking at the AFM images for the \{100\} and \{111\} orientated
single crystals of Fig. \ref{100} and \ref{111}, a clear polishing
effect can be seen. The deep nano-grooves left over from
mechanical polishing have been removed, leaving a smoother surface
over the 25 $\mu$m$^2$ image area.  The line trace perpendicular
to these polishing grooves reiterate this with a reduction in
roughness from 0.92 to 0.23 nm RMS and 0.31 to 0.09 nm RMS for the
\{100\} and \{111\} samples respectively.  After polishing the
line traces appear very similar to the traces along original
polishing direction, again showing this clear removal.  With
regards to polishing rate, these roughness values show faster
polishing on the \{100\} than the \{111\} plane as seen in
mechanical polishing.

It is worth noting that no attempt was made at polishing along
soft mechanical directions, with the sample being rotated while
sweeping across the pad.  Equal time was then spent polishing
along hard and soft polishing directions with no indication of the
fracture damage seen when mechanically polishing along hard
directions\cite{Schuelke2013}, showing the gentle nature of the
polishing mechanism used here. The surfaces are also free of
polishing debris or remaining polishing slurry due to the combined
use of SC-1 and HF cleaning after polishing, while it can also be
assumed that there has been little to no increase in subsurface
damage due to the lack of diamond grit being used in the polishing
slurry. Due the difference in properties of diamond and the
materials typically polished with CMP, study of the polishing pad
is needed to optimize the adapted technique.  Due to the hardness
of diamond additional abrasion of the pad is seen, as highlighted
in Fig. \ref{pad}C, reducing the life time of the pad.  When a
harder EP1 pad was used scratching of the diamond crystal was seen
demonstrating the importance of the pad.

It has been shown that CMP can be used efficiently to remove the
scaife marks brought about through mechanical polishing of single
crystal diamond and provide low roughness \{100\} and \{111\} diamond surfaces.
With this technique polyester/polyurethane pads are used at room
temperature with low applied pressures, without the use of diamond
grit, simplifying the process and making CMP a promising
technique.

\section{Conclusion}
\{100\} and \{111\} orientated single crystal diamond has been
polished with Chemical Mechanical Polishing through the use of a
polyester/polyurethane pad and an alkaline colloidal silica
polishing fluid.  Clear removal of the mechanical polishing
induced nano-grooves can be seen, with \{100\} surface roughness values
being reduced from 0.92 to 0.23 nm RMS along a 5 $\mu$m line trace
taken perpendicular to the direction of the nano-grooves.
Meanwhile, the equivalent line trace on the \{111\} sample shows a
reduction from 0.31 to 0.09 nm RMS.  Therefore with its simplicity
due to the use of materials commonly found in the IC fabrication
industry, along with the lack of diamond grit and elevated
temperatures, CMP is a promising technique for removing mechanical
polishing introduced artefacts.

\begin{acknowledgement}
The authors wish to thank Daniel Twitchen at Elementsix for providing
the samples used, Mark Kennedy and John McCrossan at Logitech for
their help with the early CMP process, and Rashmi Sudiwala at the
Cardiff School of Physics and Astronomy for manufacturing the
diamond holder. OAW also wishes to acknowledge the support of
Marie Curie Actions for his intra-EU fellowship.
\end{acknowledgement}

\bibliography{biblio}

\providecommand{\latin}[1]{#1}
\providecommand*\mcitethebibliography{\thebibliography}
\csname @ifundefined\endcsname{endmcitethebibliography}
  {\let\endmcitethebibliography\endthebibliography}{}
\begin{mcitethebibliography}{33}
\providecommand*\natexlab[1]{#1}
\providecommand*\mciteSetBstSublistMode[1]{}
\providecommand*\mciteSetBstMaxWidthForm[2]{}
\providecommand*\mciteBstWouldAddEndPuncttrue
  {\def\EndOfBibitem{\unskip.}}
\providecommand*\mciteBstWouldAddEndPunctfalse
  {\let\EndOfBibitem\relax}
\providecommand*\mciteSetBstMidEndSepPunct[3]{}
\providecommand*\mciteSetBstSublistLabelBeginEnd[3]{}
\providecommand*\EndOfBibitem{}
\mciteSetBstSublistMode{f}
\mciteSetBstMaxWidthForm{subitem}{(\alph{mcitesubitemcount})}
\mciteSetBstSublistLabelBeginEnd
  {\mcitemaxwidthsubitemform\space}
  {\relax}
  {\relax}

\bibitem[Field(1992)]{Field1992}
Field,~J.~E. \emph{The properties of natural and synthetic diamond}; Academic
  Press: London, 1992\relax
\mciteBstWouldAddEndPuncttrue
\mciteSetBstMidEndSepPunct{\mcitedefaultmidpunct}
{\mcitedefaultendpunct}{\mcitedefaultseppunct}\relax
\EndOfBibitem
\bibitem[Kohn \latin{et~al.}(2001)Kohn, Adamschik, Schmid, Denisenko, Aleksov,
  and Ebert]{Kohn2001}
Kohn,~E.; Adamschik,~M.; Schmid,~P.; Denisenko,~A.; Aleksov,~A.; Ebert,~W.
  \emph{Journal of Physics D: Applied Physics} \textbf{2001}, \emph{34},
  R77\relax
\mciteBstWouldAddEndPuncttrue
\mciteSetBstMidEndSepPunct{\mcitedefaultmidpunct}
{\mcitedefaultendpunct}{\mcitedefaultseppunct}\relax
\EndOfBibitem
\bibitem[Schuelke and Grotjohn(2013)Schuelke, and Grotjohn]{Schuelke2013}
Schuelke,~T.; Grotjohn,~T.~A. \emph{Diamond and Related Materials}
  \textbf{2013}, \emph{32}, 17--26\relax
\mciteBstWouldAddEndPuncttrue
\mciteSetBstMidEndSepPunct{\mcitedefaultmidpunct}
{\mcitedefaultendpunct}{\mcitedefaultseppunct}\relax
\EndOfBibitem
\bibitem[Teraji \latin{et~al.}(2004)Teraji, Yoshizaki, Wada, Hamada, and
  Ito]{Teraji2004}
Teraji,~T.; Yoshizaki,~S.; Wada,~H.; Hamada,~M.; Ito,~T. \emph{Diamond and
  Related Materials} \textbf{2004}, \emph{13}, 858--862\relax
\mciteBstWouldAddEndPuncttrue
\mciteSetBstMidEndSepPunct{\mcitedefaultmidpunct}
{\mcitedefaultendpunct}{\mcitedefaultseppunct}\relax
\EndOfBibitem
\bibitem[Isberg \latin{et~al.}(2002)Isberg, Hammersberg, Johansson, Wikström,
  Twitchen, Whitehead, Coe, and Scarsbrook]{Isberg2002}
Isberg,~J.; Hammersberg,~J.; Johansson,~E.; Wikström,~T.; Twitchen,~D.~J.;
  Whitehead,~A.~J.; Coe,~S.~E.; Scarsbrook,~G.~A. \emph{Science} \textbf{2002},
  \emph{297}, 1670--1672\relax
\mciteBstWouldAddEndPuncttrue
\mciteSetBstMidEndSepPunct{\mcitedefaultmidpunct}
{\mcitedefaultendpunct}{\mcitedefaultseppunct}\relax
\EndOfBibitem
\bibitem[Tallaire \latin{et~al.}(2005)Tallaire, Achard, Silva, Sussmann, and
  Gicquel]{Tallaire2005}
Tallaire,~A.; Achard,~J.; Silva,~F.; Sussmann,~R.~S.; Gicquel,~A. \emph{Diamond
  and Related Materials} \textbf{2005}, \emph{14}, 249--254\relax
\mciteBstWouldAddEndPuncttrue
\mciteSetBstMidEndSepPunct{\mcitedefaultmidpunct}
{\mcitedefaultendpunct}{\mcitedefaultseppunct}\relax
\EndOfBibitem
\bibitem[Friel \latin{et~al.}(2009)Friel, Clewes, Dhillon, Perkins, Twitchen,
  and Scarsbrook]{Friel2009}
Friel,~I.; Clewes,~S.~L.; Dhillon,~H.~K.; Perkins,~N.; Twitchen,~D.~J.;
  Scarsbrook,~G.~A. \emph{Diamond and Related Materials} \textbf{2009},
  \emph{18}, 808--815\relax
\mciteBstWouldAddEndPuncttrue
\mciteSetBstMidEndSepPunct{\mcitedefaultmidpunct}
{\mcitedefaultendpunct}{\mcitedefaultseppunct}\relax
\EndOfBibitem
\bibitem[Hauf \latin{et~al.}(2013)Hauf, Simon, Scifert, Holleitner, Stutzmann,
  and Garrido]{Hauf2013}
Hauf,~M.; Simon,~P.; Scifert,~M.; Holleitner,~W.; Stutzmann,~M.; Garrido,~J.
  \emph{arXiv:1310.8616v1} \textbf{2013}, \relax
\mciteBstWouldAddEndPunctfalse
\mciteSetBstMidEndSepPunct{\mcitedefaultmidpunct}
{}{\mcitedefaultseppunct}\relax
\EndOfBibitem
\bibitem[Malshe \latin{et~al.}(1999)Malshe, Park, Brown, and
  Naseem]{Malshe1999}
Malshe,~A.~P.; Park,~B.~S.; Brown,~W.~D.; Naseem,~H.~A. \emph{Diamond and
  Related Materials} \textbf{1999}, \emph{8}, 1198--1213\relax
\mciteBstWouldAddEndPuncttrue
\mciteSetBstMidEndSepPunct{\mcitedefaultmidpunct}
{\mcitedefaultendpunct}{\mcitedefaultseppunct}\relax
\EndOfBibitem
\bibitem[Jin \latin{et~al.}(1992)Jin, Graebner, Kammlott, Tiefel, Kosinski,
  Chen, and Fastnacht]{Jin1992}
Jin,~S.; Graebner,~J.~E.; Kammlott,~G.~W.; Tiefel,~T.~H.; Kosinski,~S.~G.;
  Chen,~L.~H.; Fastnacht,~R.~A. \emph{Applied Physics Letters} \textbf{1992},
  \emph{60}, 1948--1950\relax
\mciteBstWouldAddEndPuncttrue
\mciteSetBstMidEndSepPunct{\mcitedefaultmidpunct}
{\mcitedefaultendpunct}{\mcitedefaultseppunct}\relax
\EndOfBibitem
\bibitem[Suzuki \latin{et~al.}(1996)Suzuki, Yasunaga, Seki, Ide, Watanabe, and
  Uematsu]{Suzuki1996}
Suzuki,~K.; Yasunaga,~N.; Seki,~Y.; Ide,~A.; Watanabe,~N.; Uematsu,~T.
  \emph{Proceedings of ASPE} \textbf{1996}, 482--485\relax
\mciteBstWouldAddEndPuncttrue
\mciteSetBstMidEndSepPunct{\mcitedefaultmidpunct}
{\mcitedefaultendpunct}{\mcitedefaultseppunct}\relax
\EndOfBibitem
\bibitem[Hirata \latin{et~al.}(1992)Hirata, Tokura, and Yoshikawa]{Hirata1992}
Hirata,~A.; Tokura,~H.; Yoshikawa,~M. \emph{Thin Solid Films} \textbf{1992},
  \emph{212}, 43--48\relax
\mciteBstWouldAddEndPuncttrue
\mciteSetBstMidEndSepPunct{\mcitedefaultmidpunct}
{\mcitedefaultendpunct}{\mcitedefaultseppunct}\relax
\EndOfBibitem
\bibitem[Ozkan \latin{et~al.}(1997)Ozkan, Malshe, and Brown]{Ozkan1997}
Ozkan,~A.~M.; Malshe,~A.~P.; Brown,~W.~D. \emph{Diamond and Related Materials}
  \textbf{1997}, \emph{6}, 1789--1798\relax
\mciteBstWouldAddEndPuncttrue
\mciteSetBstMidEndSepPunct{\mcitedefaultmidpunct}
{\mcitedefaultendpunct}{\mcitedefaultseppunct}\relax
\EndOfBibitem
\bibitem[Olsen \latin{et~al.}(2004)Olsen, Aspinwall, and Dewes]{Olsen2004}
Olsen,~R.~H.; Aspinwall,~D.~K.; Dewes,~R.~C. \emph{Journal of Materials
  Processing Technology} \textbf{2004}, \emph{155–156}, 1227--1234\relax
\mciteBstWouldAddEndPuncttrue
\mciteSetBstMidEndSepPunct{\mcitedefaultmidpunct}
{\mcitedefaultendpunct}{\mcitedefaultseppunct}\relax
\EndOfBibitem
\bibitem[Chen and Zhang(2013)Chen, and Zhang]{Chen2013}
Chen,~Y.; Zhang,~L. \emph{Polishing of Diamond Materials}; Springer: London,
  2013; pp 25--44\relax
\mciteBstWouldAddEndPuncttrue
\mciteSetBstMidEndSepPunct{\mcitedefaultmidpunct}
{\mcitedefaultendpunct}{\mcitedefaultseppunct}\relax
\EndOfBibitem
\bibitem[Grillo and Field(1997)Grillo, and Field]{Grillo1997}
Grillo,~S.~E.; Field,~J.~E. \emph{Journal of Physics D: Applied Physics}
  \textbf{1997}, \emph{30}, 202\relax
\mciteBstWouldAddEndPuncttrue
\mciteSetBstMidEndSepPunct{\mcitedefaultmidpunct}
{\mcitedefaultendpunct}{\mcitedefaultseppunct}\relax
\EndOfBibitem
\bibitem[Pastewka \latin{et~al.}(2011)Pastewka, Moser, Gumbsch, and
  Moseler]{Pastewka2011}
Pastewka,~L.; Moser,~S.; Gumbsch,~P.; Moseler,~M. \emph{Nature Materials}
  \textbf{2011}, \emph{10}, 34--38\relax
\mciteBstWouldAddEndPuncttrue
\mciteSetBstMidEndSepPunct{\mcitedefaultmidpunct}
{\mcitedefaultendpunct}{\mcitedefaultseppunct}\relax
\EndOfBibitem
\bibitem[Hird and Field(2004)Hird, and Field]{Hird2004}
Hird,~J.~R.; Field,~J.~E. \emph{Proceedings of the Royal Society of London
  Series a-Mathematical Physical and Engineering Sciences} \textbf{2004},
  \emph{460}, 3547--3568\relax
\mciteBstWouldAddEndPuncttrue
\mciteSetBstMidEndSepPunct{\mcitedefaultmidpunct}
{\mcitedefaultendpunct}{\mcitedefaultseppunct}\relax
\EndOfBibitem
\bibitem[Huisman \latin{et~al.}(1997)Huisman, Peters, de~Vries, Vlieg, Yang,
  Derry, and van~der Veen]{Huisman1997}
Huisman,~W.~J.; Peters,~J.~F.; de~Vries,~S.~A.; Vlieg,~E.; Yang,~W.~S.;
  Derry,~T.~E.; van~der Veen,~J.~F. \emph{Surface Science} \textbf{1997},
  \emph{387}, 342--353\relax
\mciteBstWouldAddEndPuncttrue
\mciteSetBstMidEndSepPunct{\mcitedefaultmidpunct}
{\mcitedefaultendpunct}{\mcitedefaultseppunct}\relax
\EndOfBibitem
\bibitem[Volpe \latin{et~al.}(2009)Volpe, Muret, Omnes, Achard, Silva, Brinza,
  and Gicquel]{Volpe2009}
Volpe,~P.-N.; Muret,~P.; Omnes,~F.; Achard,~J.; Silva,~F.; Brinza,~O.;
  Gicquel,~A. \emph{Diamond and Related Materials} \textbf{2009}, \emph{18},
  1205--1210\relax
\mciteBstWouldAddEndPuncttrue
\mciteSetBstMidEndSepPunct{\mcitedefaultmidpunct}
{\mcitedefaultendpunct}{\mcitedefaultseppunct}\relax
\EndOfBibitem
\bibitem[Gaisinskaya \latin{et~al.}(2009)Gaisinskaya, Edrei, Hoffman, and
  Feldheim]{Gaisinskaya2009}
Gaisinskaya,~A.; Edrei,~R.; Hoffman,~A.; Feldheim,~Y. \emph{Diamond and Related
  Materials} \textbf{2009}, \emph{18}, 1466--1473\relax
\mciteBstWouldAddEndPuncttrue
\mciteSetBstMidEndSepPunct{\mcitedefaultmidpunct}
{\mcitedefaultendpunct}{\mcitedefaultseppunct}\relax
\EndOfBibitem
\bibitem[Kühnle and Weis(1995)Kühnle, and Weis]{Kuhnle1995}
Kühnle,~J.; Weis,~O. \emph{Surface Science} \textbf{1995}, \emph{340},
  16--22\relax
\mciteBstWouldAddEndPuncttrue
\mciteSetBstMidEndSepPunct{\mcitedefaultmidpunct}
{\mcitedefaultendpunct}{\mcitedefaultseppunct}\relax
\EndOfBibitem
\bibitem[Kubota \latin{et~al.}(2012)Kubota, Fukuyama, Ichimori, and
  Touge]{Kubota2012}
Kubota,~A.; Fukuyama,~S.; Ichimori,~Y.; Touge,~M. \emph{Diamond and Related
  Materials} \textbf{2012}, \emph{24}, 59--62\relax
\mciteBstWouldAddEndPuncttrue
\mciteSetBstMidEndSepPunct{\mcitedefaultmidpunct}
{\mcitedefaultendpunct}{\mcitedefaultseppunct}\relax
\EndOfBibitem
\bibitem[Wang \latin{et~al.}(2006)Wang, Zhang, Kuang, and Chen]{Wang2006}
Wang,~C.~Y.; Zhang,~F.~L.; Kuang,~T.~C.; Chen,~C.~L. \emph{Thin Solid Films}
  \textbf{2006}, \emph{496}, 698--702\relax
\mciteBstWouldAddEndPuncttrue
\mciteSetBstMidEndSepPunct{\mcitedefaultmidpunct}
{\mcitedefaultendpunct}{\mcitedefaultseppunct}\relax
\EndOfBibitem
\bibitem[Haisma \latin{et~al.}(1992)Haisma, Vanderkruis, Spierings, Oomen, and
  Fey]{Haisma1992}
Haisma,~J.; Vanderkruis,~F.; Spierings,~B.; Oomen,~J.~M.; Fey,~F.
  \emph{Precision Engineering-Journal of the American Society for Precision
  Engineering} \textbf{1992}, \emph{14}, 20--27\relax
\mciteBstWouldAddEndPuncttrue
\mciteSetBstMidEndSepPunct{\mcitedefaultmidpunct}
{\mcitedefaultendpunct}{\mcitedefaultseppunct}\relax
\EndOfBibitem
\bibitem[Krishnan \latin{et~al.}(2009)Krishnan, Nalaskowski, and
  Cook]{Krishnan2009}
Krishnan,~M.; Nalaskowski,~J.~W.; Cook,~L.~M. \emph{Chemical Reviews}
  \textbf{2009}, \emph{110}, 178--204\relax
\mciteBstWouldAddEndPuncttrue
\mciteSetBstMidEndSepPunct{\mcitedefaultmidpunct}
{\mcitedefaultendpunct}{\mcitedefaultseppunct}\relax
\EndOfBibitem
\bibitem[Thomas \latin{et~al.}(2014)Thomas, Nelson, Mandal, Foord, and
  Williams]{Thomas2014}
Thomas,~E. L.~H.; Nelson,~G.~W.; Mandal,~S.; Foord,~J.~S.; Williams,~O.~A.
  \emph{Carbon} \textbf{2014}, \emph{68}, 473--479\relax
\mciteBstWouldAddEndPuncttrue
\mciteSetBstMidEndSepPunct{\mcitedefaultmidpunct}
{\mcitedefaultendpunct}{\mcitedefaultseppunct}\relax
\EndOfBibitem
\bibitem[Hocheng \latin{et~al.}(2001)Hocheng, Tsai, and Su]{Hocheng2001}
Hocheng,~H.; Tsai,~H.~Y.; Su,~Y.~T. \emph{Journal of the Electrochemical
  Society} \textbf{2001}, \emph{148}, G581--G586\relax
\mciteBstWouldAddEndPuncttrue
\mciteSetBstMidEndSepPunct{\mcitedefaultmidpunct}
{\mcitedefaultendpunct}{\mcitedefaultseppunct}\relax
\EndOfBibitem
\bibitem[Hooper \latin{et~al.}(2002)Hooper, Byrne, and Galligan]{Hooper2002}
Hooper,~B.~J.; Byrne,~G.; Galligan,~S. \emph{Journal of Materials Processing
  Technology} \textbf{2002}, \emph{123}, 107--113\relax
\mciteBstWouldAddEndPuncttrue
\mciteSetBstMidEndSepPunct{\mcitedefaultmidpunct}
{\mcitedefaultendpunct}{\mcitedefaultseppunct}\relax
\EndOfBibitem
\bibitem[Lu \latin{et~al.}(2002)Lu, Fookes, Obeng, Machinski, and
  Richardson]{Lu2002}
Lu,~H.; Fookes,~B.; Obeng,~Y.; Machinski,~S.; Richardson,~K.~A. \emph{Materials
  Characterization} \textbf{2002}, \emph{49}, 35--44\relax
\mciteBstWouldAddEndPuncttrue
\mciteSetBstMidEndSepPunct{\mcitedefaultmidpunct}
{\mcitedefaultendpunct}{\mcitedefaultseppunct}\relax
\EndOfBibitem
\bibitem[McGrath and Davis(2004)McGrath, and Davis]{Mcgrath2004}
McGrath,~J.; Davis,~C. \emph{Journal of Materials Processing Technology}
  \textbf{2004}, \emph{153}, 666--673\relax
\mciteBstWouldAddEndPuncttrue
\mciteSetBstMidEndSepPunct{\mcitedefaultmidpunct}
{\mcitedefaultendpunct}{\mcitedefaultseppunct}\relax
\EndOfBibitem
\bibitem[Park \latin{et~al.}(2007)Park, Kim, Chang, and Jeong]{Park2007}
Park,~K.~H.; Kim,~H.~J.; Chang,~O.~M.; Jeong,~H.~D. \emph{Journal of Materials
  Processing Technology} \textbf{2007}, \emph{187}, 73--76\relax
\mciteBstWouldAddEndPuncttrue
\mciteSetBstMidEndSepPunct{\mcitedefaultmidpunct}
{\mcitedefaultendpunct}{\mcitedefaultseppunct}\relax
\EndOfBibitem
\end{mcitethebibliography}

\end{document}